\documentclass[12pt]{article}
\usepackage{graphicx}
\usepackage{epsf}

\makeatletter
\def\@biblabel#1{}
\makeatother

\begin{document}



\noindent
{\bf SN 2011ht: WEAK EXPLOSION IN MASSIVE EXTENDED \\
ENVELOPE}

\bigskip

N. N. Chugai

\bigskip

{\it Institute of Astronomy of Russian Academy of Sciences, Moscow}\\


\vspace{1cm}
\begin{abstract}
A possibility is explored to account for the light curve and the low expansion 
velocity of the supernova SN~2011ht, a member of group of three objects 
showing signatures of both IIn and IIP supernovae. 
It is argued that the radiated energy and the expansion velocity are consistent 
with the low energy explosion ($\approx6\times10^{49}$ erg) and $\leq2~M_{\odot}$
 ejecta interacting with the circumstellar envelope of $6-8~M_{\odot}$ 
and the radius of $\sim2\times10^{14}$ cm. The test of this scenario is 
proposed.

\end{abstract}


\vspace{1cm}

\section{Introduction}

Among type IIn supernovae (SN~IIn, with "n" standing for "narrow lines") that are commonly associated 
with the presence of a dense circumstellar medium there is a unique variety 
composed of SN~1994W (Sollerman et al. 1998), SN~2009kn (Kankare et al. 2012), 
and SN~2011ht (Roming et al. 2012; Mauerhan et al. 2013). Their bolometric light 
curve has $\sim120$ days plateau reminiscent of SN~IIP. The plateau ends up with the 
luminosity drop by a factor of ten and a subsequent exit to the 
tail somewhat similar to the radioactive tail of SN~IIP but probably of different 
origin. Maximum with $\sim-18$ mag is attained at about day 40. The spectrum 
is smooth continuum with strong emission lines of  H$\alpha$ and H$\beta$ 
characterized by the narrow core (FWHM$\sim 700-800$ km s$^{-1}$) and 
broad wings $\sim \pm5000$ km s$^{-1}$. Apart from hydrogen lines the 
spectrum shows narrow metal lines, mostly Fe\,II, with velocity of absorption 
minima of $\sim-(500...~700)$ km s$^{-1}$. The similarity of light curves and 
spectra of the mentioned supernovae justifies their selection into a 
special group designated SN~IIn-P (Mauerhan et al. 2013); the notation 
emphasises their resemblance with SN~IIn and SN~IIP first mentioned for 
SN~1994W (Sollerman et al. 1998). 

The model of SN~1994W proposed earlier suggested the explosion of the 
red supergiant with $7~M_{\odot}$ ejecta and the kinetic energy 
of $\sim10^{51}$ erg (Chugai et al. 2004). According to this scenario 
(dubbed as scenario A) the supernova interacts with the dense extended 
circumstellar (CS) envelope; narrow lines form in the CS envelope 
expanded at $\sim10^3$ km s$^{-1}$; broad wings are produced by the scattering of
line photons on thermal electrons of the same CS envelope. This scenario, 
however, faces a serious problem, because it is becoming clear that at 
the late stage ($t>120$ d) supernovae SN~IIn-P do not show signatures of the 
high-velocity material ($\sim4000$ km s$^{-1}$) that is predicted by this scenario. 
One might suggest that this gas is not seen because the cool dense shell (CDS)
at the contact surface between supernova and CS material is very opaque. 
At the late stage this situation might occur, if the dust forms in the CDS. 
Yet the case of SN~1998S where the dust indeed seem to form in the CDS 
(Pozzo et al. 2004) broad emission lines are seen, possibly because of the 
mixing of the fragments of the CDS with the hot gas of the forward 
shock.

In the alternative scenario (call it B) proposed by Dessart et al. (2009) 
the  spectrum of SN~1994W, including the continuum and lines, 
forms in a massive envelope with low expansion velocity ($\sim1000$ km s$^{-1}$) 
implied by narrow lines. In fact, authors have demonstrated that the expanding atmosphere 
with a steep density gradient, the effective temperature of $\sim7000$~K, and 
photosphere radius of $\sim10^{15}$ cm reproduces the observed spectrum fairly 
well. The success of the straightforward scenario in the modelling of the non-trivial spectrum 
makes this scenario  very attractive. Noteworthy, the scanario B does not 
contain high velocity gas unlike the scenario A.

The scenario B, however, leaves open a question, whether the 
energy requirements are consistent with the low expansion velocity. 
The present paper is focused on this issue.  To this end a model is developed 
to describe the phenomenon of SN~2011ht for which 
most complete observations are available compared to other two SN~IIn-P. 
The model is based on the thin shell approximation that is commonly used 
for the analysis of SN~IIn. Here, however, the model includes  diffusion of the 
trapped radiation in the optically thick envelope. The section 2 describes the 
model, while the section 3 presents results of the light curve modelling. 
The modelling of line profiles of H$\alpha$ and H$\gamma$ is presented in section 4. 
Note, the simultaneous description of these lines in the framework of the 
unified model of the emission and Thomson scattering in CS envelope turned out 
problematic in the former scenario of SN~1994W (Chugai et al. 2004).

The discovery on 2011 September 29 (JD=2455834) caught SN~2011ht during the 
rapid flux rise (Roming et al. 2012; Mauerhan et al. 2013). 
It is reasonable to admit, therefore, that the explosion took place a few days before the discovery. Here the explosion date JD=2455830 is adopted.

\section{General considerations and model}

The velocity at the photosphere of SN~2011ht fixed by absorption minima, 
e.g., H$\alpha$ is about 600 km s$^{-1}$ (Mauerhan et al. 2013); this  
value remains constant through the spectral observations ($t>30$ d) at the plateau 
stage. The latter indicates that the velocity dispersion and the relative thickness 
of the shell are rather small. Furthermore, the velocity persistence also sugggests 
that the shell acceleration phase is brief, $t_a\leq30$ d, which means in turn 
that the external radius of the CS envelope is  
$R_{cs}\sim vt_a\leq 2\times10^{14}$ cm. The main stage of the radiative cooling 
($\sim120$ d) therefore should be considered as the result of slow diffusion 
of the trapped 
radiation generated at the early phase $t\leq 30$ d. In this respect SN~2011ht 
is similar to SN~IIP. The difference is that the bulk of SN~IIP matter is distributed 
in a wide range of velocities which is manifested in the significant decrease 
of the photospheric velocity at the plateau in contrast to SN~IIn-P.
The diffusion time characterized by the plateau stage is several times greater 
than the acceleration time, so a significant fraction of the internal 
energy is spent on the pressure work at the plateau stage.
The strong raiation-dominated shock in the uniform medium deposits 80\% of the 
energy in the internal energy (i.e., radiation) and 20\% in the kinetic energy 
(Chevalier 1976). Assuming that the initial internal energy is equally shared between 
the work on the expansion and the escaped radiation, we conclude that roughly 2/3 of the 
explosion energy is spent on the kinetic energy of the accelerated shell, while 
1/3 is escaped radiation. Given the radiated energy of SN~2011ht of $\approx 2\times10^{49}$ erg 
(Mauerhan et al. 2013) we thus conclude that the kinetic energy is 
 $\sim 4\times10^{49}$ erg and the explosion energy is $E\sim6\times10^{49}$ erg.
Taking into account the expansion velocity of $v\approx600$ km s$^{-1}$ and the estimated kinetic energy we infer the total mass of the expanding shell 
as $M\sim10~M_{\odot}$.


\begin{table}[t]

\vspace{6mm}
\centering
{{\bf Table:} Parameters of SN~2011ht models}
\label{t-par} 

\vspace{5mm}\begin{tabular}{l|c|c|c|c|c|c} \hline\hline
Model & $E$ & $M_{sn}$ & $M_{cs}$ & $R_{cs}$  & s & $E_r$\\
  & $10^{50}$ erg & {$M_{\odot}$}& {$M_{\odot}$} & $10^{14}$ cm & &  $10^{50}$ erg\\
 \hline
   m1  & 0.6   & 0.01  &   10  &  2   &   0  &  1.9\\
   m2  & 0.6   &  2    &   8   &  2   &   0  &  1.5\\ 
   m3  & 0.57  &  2    &   6.5 &  2.5 &   0  &  1.7\\
   m3f & 0.57  &  2    &   6.5 &  2.5 &   0  &  2.5\\ 
   m4  & 0.6   &  2    &   7   &  3.5 &   2  &  2.0\\ 
   m4f & 0.6   &  2    &   7   &  3.5 &   2  &  3.0\\ 
\hline

\end{tabular}
\end{table}

The small relative thickness of the shell and the brief acceleration phase
prompts us a simple model based on the thin shell approximation (Giuliani 1982).
We consider geometrically thin, but optically thick, shell with the interior 
filled by the radiation. The radiation energy is determined by the dissipation 
of the kinetic energy of the supernova ejecta, by the work of 
the radiation pressure, and the radiation escape via diffusion.
The equation of motion for this shell is (cf. Giuliani 1982)
\begin{equation}
M\frac{dv}{dt} = 4\pi R^2\left[\rho_{sn}\left(\frac{R}{t}-v\right)^2+
p-\rho_{cs}v^2\right]\,,
\label{eq-mom}
\end{equation}
where $M$ is the mass of the shell with the radius $R$, $v$ is the shell expansion velocity,
 $\rho_{sn}$ is the supernova density at the radius $R$, 
 $\rho_{cs}$ is the CS density at the radius $R$,  and 
$p$ is the radiation pressure that is assumed to be uniform in the cavity. 
The expansion velocity of undisturbed CS matter is presumably negligibly small.
Undisturbed supernova ejecta are assumed to expand homologously, 
($v=r/t$) with the density distribution  
$\rho \propto (v_0/v)/(1+(v/v_0)^7)$, i.e., $\rho \propto v^{-1}$ in the inner 
zone ($v<v_0$) and $\propto v^{-8}$ in the outer layers.

The radiation pressure in the cavity is $p=E_r/(4\pi R^3)$, where 
$E_r$ is the radiation energy described by the equation
\begin{equation}
\frac{dE_r}{dt} = 2\pi R^2\rho_{sn}\left(\frac{R}{t}-v\right)^3
-E_r\frac{v}{R}-\frac{E_r}{t_c}\,.
\label{eq-ener}
\end{equation}
The first term in the right hand side is the rate of the internal energy generation 
due to the ejecta collision  with the thin shell, the second term is the 
work of the radiation pressure, and the last term is the luminosity due to 
the radiation diffusion. The luminosity is determined as $L=E_r/t_d$, 
where the diffusion time is 
\begin{equation}
t_d=\xi\frac{R}{c}\tau\,.
\label{eq-tcool}
\end{equation}
Here $\tau\gg1$ is the shell optical depth, $c$ is the speed of light, 
and  $\xi$ is a factor of order unity related to the geometry. For the central 
source in the uniform sphere $\xi=0.5$ (Sunyaev and Titarchuk 1980); a similar 
value one obtains for the geometrically thin shell filled by the isotropic radiation. We adopt $\xi=0.5$. 

The shell optical depth is calculated using Rosseland opacity (Alexander 1975). 
The temperature distribution in the shell is 
determined iteratively on the bases of the Eddington solution for the 
plane slab, $T^4=(3/4)T_e^4(2/3+\tau)$, where $T_e$ is the effective temperature.
The shell density is assumed to be equal $\rho_s=7\rho_{cs}$ in line with 
the density jump in the strong radiation-dominated shock.
After the shock break out of the CS envelope ($R>R_{cs}$) the shell density 
is set to be $\rho_s\propto (R_{cs}/R)^3$ implied by the free expansion. 
The system of equations of motion, energy, and mass conservation is solved by 
Runge-Kutta of 4-th order. In every case the energy is conserved with the 
accuracy of 1\%. 

To test the model we calculated the result of a
central explosive release of $10^{51}$ erg in the uniform envelope of 
$4.2~M_{\odot}$ and the radius of $5\times10^{13}$ cm. The explosion is 
simulated by the kinetic energy of $0.01~M_{\odot}$ shell. In this  
formulation the modelling in the framework of 
the radiation hydrodynamic is available 
(Chevalier 1976, model A). The light curve in our model slightly differs 
from that of the model A but the length of the plateau in both models 
(57 days) coinside within one day. Despite its 
simplicity our model thus catches the essence of the acceleration dynamics and 
the light curve produced by the explosion in an extended envelope.

\begin{figure}[h]
\centering
\includegraphics[width=0.9\textwidth]{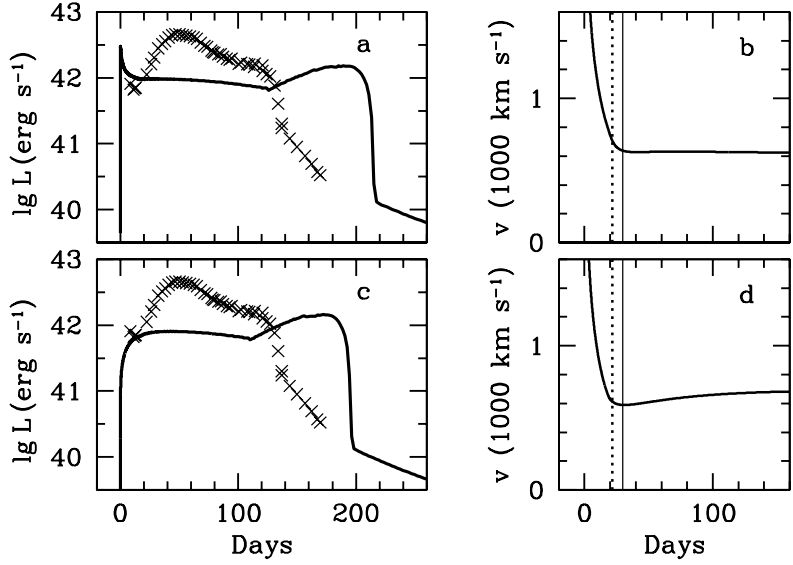}
\vspace{-8mm}
\caption{\rm \footnotesize 
Model bolometric light curve (panels {\it a} and {\it c}), compared to 
observations of SN~2011ht ({\it crosses}, Mauerhan et al. 2013; Roming et al. 2012), 
and the model thin shell velocity (panels {\it b} and {\it d}). Model m1 (cf. Table) 
is ploted in panels {\it a} and {\it b}, model m2 is in panels 
{\it c} and {\it d}. The vertical {\it solid} line in panels {\it b} and {\it d} 
corresponds to the epoch of the first spectrum (day 30), while the {\it dotted} line 
shows the moment when the thin shell radius is equal to $R_{cs}$. 
}
\label{f-lc1}
\end{figure}

\section{Modelling results} 

SN~2011ht parameter estimates recovered above are used here for two illustrative 
models, m1 and m2 (cf. Table and Fig.1). The Table contains the explosion energy and  
mass of SN ejecta, the mass of the CS envelope, its radius $R_{cs}$, the 
power index $s$ of the density distribution, $\rho\propto r^{-s}$, in the range of
$r<R_{cs}$, and the radiated energy. In both models $s=0$ while 
$\rho\propto r^{-6}$ for $r>R_{cs}$. The model m1 with the ejecta mass of 
$0.01~M_{\odot}$ in fact simulates the central explosion since the kinetic energy 
of the low mass ejecta is rapidly ($t<1$ d) thermalized. The aggregated mass of 
the supernova ejecta and CS envelope is $10~M_{\odot}$ in both models.
The models sensibly reproduce the total radiated energy (cf. Table) and the 
observed expansion velocity of SN~2011ht after day 30. Both models however have 
apparent drawbacks: the plateau is uacceptably long and the shape of 
the light curve is unlike the observed one. Particularly, the model does not show 
the hump at about day 50. Note, the late hump of the model light curve at the 
plateau end is related to the sharp drop of the opacity with the 
temperature decrease around $10^4$~K. 

The comparison of our simplified model with hydrodynamic simulation in the 
previous section suggests that the strong disagrement between the model and 
observed plateau duration is unlikely, although cannot be rulled out completely. 
The problem with the light curve description might arise because some 
relevant physics is not included in our model. 
We admit that the missing factor is the fragmentation of the swept-up shell 
as a result of either the Rayleigh-Taylor instability arising from the 
rapid deceleration of the shell (Fig.1), or the thin shell instability 
in the case of the radiative forward shock (Vishniac 1983). The outcome of the fragmentation is the decrease of the shell effective optical depth and, 
as a result, rapid radiation diffusion and larger luminosity at the early 
epoch. 

The fragmentation effect in the light curve can be implemented 
using the following description. A homogeneous spherical 
layer with the optical depth $\tau$ breaks down into spherical fragments 
of a radius $a$ with $N$ to be the number density of fragments. The gas density 
in fragments is assumed to be the same as in the smooth shell, while the intercloud density to be negligibly small. The average number of clouds along the shell radius 
is then $\tau_{oc}=\pi a^2N\Delta r$, where $\Delta r$ is the shell thickness.
Using $\tau_{oc}$ one can write the expression for the effective optical depth 
of the cloudy shell (cf. Chugai \& Chevalier 2005) as
\begin{equation}
\tau_{eff}=\tau_{oc}[1-\exp{(-\tau/\tau_{oc})}]\,.
\label{eq-tauef}
\end{equation}
This expression deviates from the exact one (Utrobin \& Chugai 2015) by less 
than 3\%. In the limit of $\tau_{oc}\gg\tau$ the relation (4) 
reproduces the optical depth of the smooth shell, $\tau_{eff}=\tau$, while for $\tau_{oc}\ll\tau$ the 
effective optical depth is reduced to average number of clouds along the shell 
radius, $\tau_{eff}=\tau_{oc}$. The fragmentation evolution is described via the time dependence of $\tau_{oc}$ 
\begin{equation}
\tau_{oc}=\tau_{oc,2}+\tau_{oc,1}/[1+(t/t_f)^6]\,.
\end{equation}
The value of $\tau_{oc,1}$ is set to meet the requirement $\tau_{oc,1}\gg\tau_0$, 
where $\tau_0$ is the initial optical depth of the smooth shell. We assume 
$\tau_{oc,1}=5\tau_0$ and $\tau_{oc,2}=200$: the choice guaratees that at the 
early epoch $\tau_{eff}$ is equal to the optical depth of the smooth shell. 
This description suggests that the fragmentation becomes significant at the 
stage $t\geq t_f$.

\begin{figure}[h]
\centering
\includegraphics[width=0.9\textwidth]{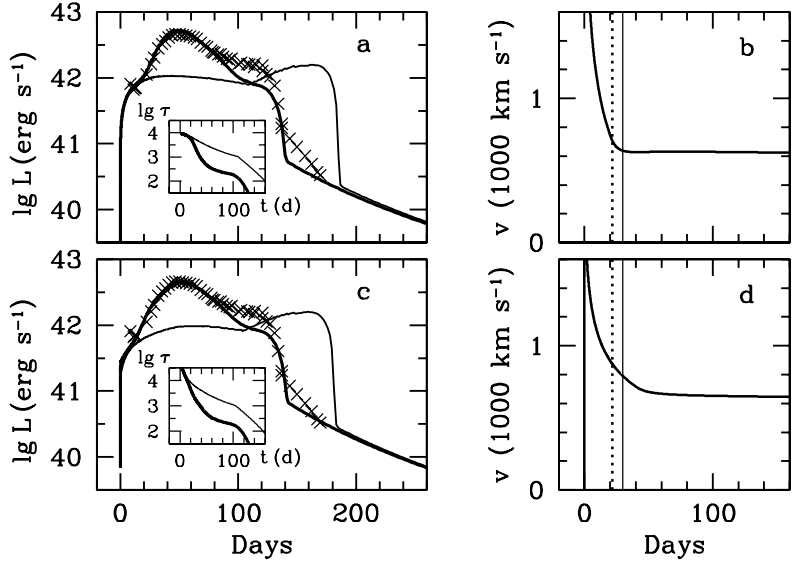}
\vspace{-8mm}
\caption{\rm \footnotesize 
The same as Fig.1, but for models with the fragmentation (m3f and m4f, cf. Table).
{\it Thin} line shows the model without fragmentation.
{\it Insets} in panels {\it a} and {\it c} show evolution of the Rosseland optical 
depth in the model without fragmentation ({\it thin} line) and with 
fragmentation.
}
\label{f-lc2}
\end{figure}

The fragmentation effect in the light curve is demonstrated by models 
m3f and m4f (Table, Fig.2) in comparison with models m3 and m4 without 
fragmentation. Models m3f and m4f differ by the intitial density distribution 
of the CS envelope: in the model m3f the density is uniform ($s=0$), while 
in the model m4f $s=2$. The fragmentation time is $t_f=13$~d in the model m3f and 
10 days in the model m4f. Model light curves describe better principal 
features of the observed light curve: maximum at about day 50 and the plateau 
duration. The shell expansion velocity in both models is close to 600 km s$^{-1}$, 
in accord with observations. Note that the velocity exit on the constant regime in the model m4f occurs later than both in the model m3f and observations; we therefore  
conclude that the model m3f is 
preferred. Exercises with different ejecta mass confirm an obvious 
guess: a model with larger mass exits to the constant velocity later because of  
the larger momentum. The model m3f exits to the constant velocity at about day 30, 
and thus demonstrates that the ejecta mass cannot exceed significantly $2~M_{\odot}$.
At the luminosity maximum ($\approx 50$ d) the effective temperature in the model 
m3f is $10^4$~K in agreement with the temperature inferred from the spectral energy 
distribution (Mauerhan et al. 2013). This additionally lends credibility that 
the model reflects basic physics of SN~2011ht phenomenon.

For the model verification of great interest is the early stage $t<t_a$ that 
preceeds complete sweeping of the CS envelope, i.e., when the thin shell radius 
$R<R_{cs}$. In the model m3f this stage corresponds to $t<t_a=21.6$ d (Fig.2). 
Our scenario predicts that at $t<t_a$ the photosphere radius is constant and equal to $R_{cs}$, while the velocity at the photosphere should coinside with the velocity of the 
undisturbed CS envelope, which is presumably small. We expect in this case that 
the spectrum at $t<t_a$ with the resolution $>100$ km s$^{-1}$ will 
not reveal absorption lines, while core of emission hydrogen lines will 
be narrower than at the late time, $t>t_a$.

Summing up, the radiated energy and the low expansion velocity of SN~2011ht are consistent with the explosion of supernova of low energy ($\approx6\times10^{49}$ erg) in the CS envelope of the radius  $\sim2\times10^{14}$ cm with the total mass of 
the swept-up shell of $\approx8-9~M_{\odot}$.

\section{Hydrogen line profiles}

In the proposed scenario the shock wave sweeps up the CS envelope in the initial 
20-30 days. After that the shell expands freely with the velocity of 
$\approx600$ km s$^{-1}$ and kinematics of $v=r/t$. The hydrogen line emission 
produced by the shell with the large Thomson optical depth, and the hydrogen 
absorption arising 
from the external rarefied layer can provide us with an additional test of the SN~2011ht model.
The Monte Carlo technique is used below to model hydrogen line profiles. 

Compared to the similar modelling of H$\alpha$ formed in the cocoon of SN~1998S with the Thomson optical depth of $\tau_T \approx 3$ (Chugai 2001), in the case of SN~2011ht 
the optical depth of the emitting shell is tremendous ($\tau_T> 10^2$), so 
one has to take into account the true absorption of quanta between scatterings.
This process is modelled by introducing the absorption probability $p=k_a/(k_a+k_T)$, where  $k_a$ and $k_T$ are the coefficients of absorption and Thomson scattering 
respectively. It is assumed that the shell consists of two components: 
optically thick ($\tau_T >10^2$) spherical layer in the velocity range of 
$v_1<v<v_2$ and the optically thin ($\tau_T =0.5$) external layer in the 
velocity range $v_2<v<v_3$ responsible for the absorption component. The ratio 
of line and continuum emission coefficients is assumed to be constant in the 
envelope; the electron temperature is set to be $10^4$~K.

As an example we consider the spectrum of SN~2011ht on day 37  
(Mauerhan et al. 2013). According to the model m3f at this stage the shell optical depth is $\tau_T \approx 10^3$. We adopt $\tau_T = 10^3$ and note that the 
variation of this value in the range of factor three does not affect the 
result significantly. The resonance optical depth in lines is assumed to be 
very large in the range of $v_1<v<v_3$. Computations of the H$\alpha$ profile 
for the extended set of parameters led us to the optimal choice 
$v_1=550$ km s$^{-1}$, $v_2=650$ km s$^{-1}$, $v_3=850$ km s$^{-1}$, and 
$p=0.09$ (Fig.3). To model H$\gamma$ one has to take into account that the
absorption probability should be smaller than in the H$\alpha$ band because 
the absorption is determined primarily by the Paschen continuum. 
If one takes into account only this absorption mechanism then for the H$\gamma$ band 
one gets $p=0.026$.  However, the H$\gamma$ calculated with this value shows 
very strong broad wings due to large number of scatterings on thermal electrons. 
The best agreement with the observed spectrum is found for $p=0.054$. 
The absorption coefficient is larger than the Paschen value possibly because of the 
contribution of numerous metall lines in this band. With the correction for the uncertainty of $p$ for H$\gamma$, we find that both H$\alpha$ and H$\gamma$ are well described 
by the unified model of the SN~2011ht consistent with the light curve model 
as regards principal parameters (velocity, temperature, and optical depth).
This success demonstrates advantage 
of the scenario B over scenario A: in the latter H$\gamma$ profile could not 
be reproduced solely by the emission and scattering in the undisturbed 
CS envelope (Chugai et al. 2004). 

\begin{figure}[h]
\centering
\includegraphics[width=0.9\textwidth]{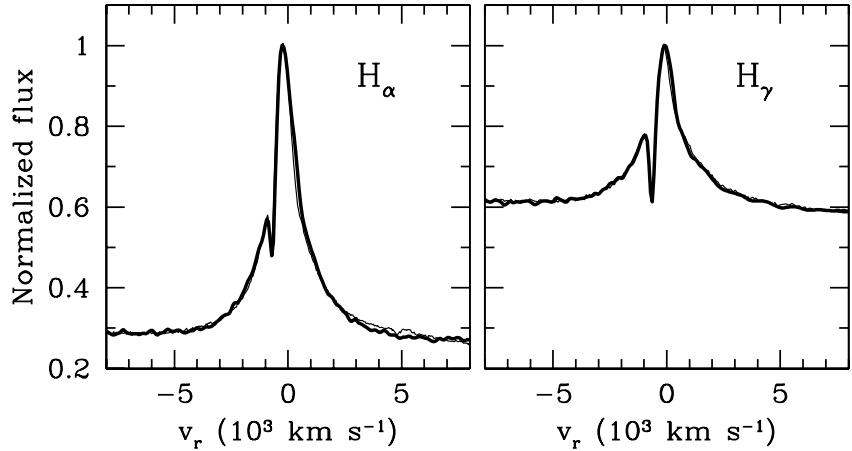}
\vspace{5mm}
\caption{\rm \footnotesize 
Model profiles of H$\alpha$ and H$\gamma$ ({\it thick} line) compared to 
the observed spectrum (Mauerhan et al. 2013). Small excess of the observed flux in the 
red wing of H$\alpha$ at about 5000 km s$^{-1}$ is caused by the presence of weak 
He\,I 6678 \AA\ line.
}
\label{f-sp}
\end{figure}

\section{Discussion and Conclusions}

The goal of the paper was the answer to the question, whether the light curve 
and spectra of supernovae IIn-P were consistent with the scenario B prompted 
by the idea of Dessart et al. (2009) that the spectrum of SN~1994W 
formed in the slowly expanding envelope ($<1000$ km s$^{-1}$). 
In the thin shell approximation with radiative diffusion  
a simple model was developed to compute the luminosity and dynamics of SN~2011ht.
The modelling demonstrates that the light curve and the low expansion velocity 
are consistent with the low energy explosion ($\approx 6\times10^{49}$ erg) and 
ejected mass $\leq2~M_{\odot}$ occured in the
CS envelope with the radius of $\sim2\times10^{14}$ cm and the mass of 
 $6-8~M_{\odot}$. In this scenario a better agreement with the observed 
light curve is achieved, if one admits the shell fragmentation.
The issue of instabilities that give rise to the fragmentation is beyond the  
scope of the present paper. It should be emphasised also that we cannot rule out 
that in the framework of the radiation hydrodynamics
one will be able to reproduce all the observations without invoking fragmentation.

The scenario B of the low energy explosion 
applied to SN~2011ht notably differ from the scenario A proposed for SN~1994W 
(Chugai et al. 2004). Major differencies of the new scenario from the old 
one are: (i) factor ten lower energy and, as a result, the lower expansion 
velocity ($<1000$ km s$^{-1}$), (ii) factor ten smaller radius of the 
CS envelope, and last but not least (iii) the line emitting region in the 
new scenario is the massive ($\sim 8~M_{\odot}$) shell accelerated by supernova explosion, while in the old scenario it was the undisturbed CS envelope with the 
mass of $\sim0.5~M_{\odot}$. The scenario B is favourable by two reasons. 
First, late time  spectra ($t>120$ d) of SN~IIn-P do not show high expansion 
velocities which is a serious problem for the scenario A. Second, the emission of 
hydrogen lines by the low velocity optically thick shell permits one to describe 
all the hydrogen lines as demonstrated by the modelling of the 
H$\alpha$ and H$\gamma$ lines. In contrast, in the scenario A the 
H$\gamma$ and H$\alpha$ lines cannot be reproduced simultaneously in the 
model of the emitting undisturbed CS envelope.
The concept of the low energy explosion for SN~IIn-P events has been 
proposed earlier by Smith (2013); he attributes this subclass along 
with the SN~1054 (Crab) to the electron-capture supernovae.

Remarkably, the scenario B admits an observational test.
It is based on the prediction that at the early stage preceeding the total 
acceleration of the CS envelope, i.e., at $t<t_a\sim 20$ days, the radial 
velocities of line absorptions should be equal to the expansion velocity 
of the undisturbed 
CS envelope, while the core of hydrogen emission lines should be significantly 
narrower than at the later epoch ($t>t_a$). 

The genesis of SN~IIn-P is an open issue. Sollerman et al. (1998) have 
mentioned two possibilities for SN~1994W: star with the initial mass from the range  
of $8-10~M_{\odot}$ with a core collapsing to the neutron star, or massive 
star ($M\geq 25~M_{\odot}$) leaving behind the black hole. Both scenario 
account for the absence of large amount of ejected $^{56}$Ni. 
From the point of view of producing a close massive CS envelope the progenitor
with the mass of  $\sim10~M_{\odot}$ is preferred. Indeed, several 
years prior to the SN outburst the explosive flash of degenerate neon may 
result in the ejection all the pre-SN envelope (Woosley et al. 2002); for the 
massive star $>25~M_{\odot}$ that heavy mass loss prior to the collapse is 
unlikely because the final nuclear burning occurs in non-degenerate fashion.
In the case of $\sim10~M_{\odot}$ progenitor the CS envelope with the radius of 
$2\times10^{14}$ cm forms by 3 yr prior to collapse provided the mass outflow velocity 
is 20 km s$^{-1}$. Notably, the low explosion energy of SN~2011ht 
($\sim 6\times10^{49}$ erg) is consistent with the prediction of neutrino mechanism 
for $\approx10~M_{\odot}$ progenitor (Kitaura et al. 2006). 
Yet, it is noteworthy, 
that the collapse of massive star ($>25~M_{\odot}$) also can produce weak explosion 
(Woosley et al. 2002).

\vspace{1cm}
I am grateful to Jon Mauerhan for the spectra of SN~2011ht.

\pagebreak   


\end{document}